\documentclass[conference]{IEEEtran}

\usepackage{cite}
\usepackage{amsmath,amssymb,amsfonts}
\usepackage{algorithmic}
\usepackage{graphicx}
\usepackage{textcomp}
\usepackage{xcolor}
\usepackage{multirow}
\usepackage{amsthm}
\usepackage{array}
\usepackage{makecell}
\usepackage[T1]{fontenc}
\usepackage{verbatim}
\usepackage[left=1.62cm,right=1.62cm,top=1.9cm]{geometry}

\newcolumntype{P}[1]{>{\centering\arraybackslash}p{#1}}

\newtheorem{theorem}{Theorem}
\newtheorem{lemma}{Lemma}

\def\BibTeX{{\rm B\kern-.05em{\sc i\kern-.025em b}\kern-.08em
		T\kern-.1667em\lower.7ex\hbox{E}\kern-.125emX}}
\begin{document}
	
	\title{OpTree: An Efficient Algorithm for All-gather Operation in Optical Interconnect Systems \\
	}
	%NFR: Efficient Communication Scheme for Distributed Model-parallel DNN Training on Optical Interconnect System
	%\author*[1]{\fnm{Fei} \sur{Dai}}\email{travis@cs.otago.ac.nz}
	
	%\author[1]{\fnm{Yawen} \sur{Chen}}\email{yawen@cs.otago.ac.nz}
	
	%\author[1]{\fnm{Zhiyi} \sur{Huang}}\email{hzy@cs.otago.ac.nz}
	
	%\author[1]{\fnm{Haibo} \sur{Zhang}}\email{haibo@cs.otago.ac.nz}
	
	%\affil*[1]{\orgdiv{Department of Computer Science}, \orgname{University of Otago}, \orgaddress{\city{Dunedin}, \postcode{9054}, \state{Otago}, \country{New Zealand}}}
	%\IEEEauthorrefmark{1}
	% \thanks{corresponding author}\footnote{Fei Dai is the corresponding author.},
	\author{\IEEEauthorblockN{Fei Dai$^*$,
			Yawen Chen,
			Zhiyi Huang, 
			Haibo Zhang 
		}
		\IEEEauthorblockA{ Department of Computer Science,
			University of Otago, Dunedin, New Zealand}
		%\IEEEauthorblockA{\IEEEauthorrefmark{2}Department of Electrical Engineering and Automation, Qilu University of Technology, Jinan, China}% <-this % stops an unwanted space
		Email:\{travis, yawen, hzy, haibo\}@cs.otago.ac.nz
	}
	%$^{(1)}$
	%\thanks{Manuscript received December 1, 2012; revised August 26, 2015. Corresponding author: M. Shell (email: http://www.michaelshell.org/contact.html).}

	%\author{\IEEEauthorblockN{1\textsuperscript{st} Fei Dai}
	%\IEEEauthorblockA{\textit{Department of Computer Science} \\
	%\textit{University of Otago}\\
	%Dunedin, New Zealand \\
	%travis@cs.otago.ac.nz}
	%\and
	%\IEEEauthorblockN{2\textsuperscript{nd} Yawen Chen}
	%\IEEEauthorblockA{\textit{Department of Computer Science} \\
	%\textit{University of Otago}\\
	%Dunedin, New Zealand \\
	%yawen@cs.otago.ac.nz}
	%\and
	%\IEEEauthorblockN{3\textsuperscript{rd} Zhiyi Huang}
	%\IEEEauthorblockA{\textit{Department of Computer Science} \\
	%\textit{University of Otago}\\
	%Dunedin, New Zealand \\
	%zhiyi@cs.otago.ac.nz}
	%\and
	%\IEEEauthorblockN{4\textsuperscript{th} Haibo Zhang}
	%\IEEEauthorblockA{\textit{Department of Computer Science} \\
	%\textit{University of Otago}\\
	%Dunedin, New Zealand \\
	%haibo@cs.otago.ac.nz}
	%\and
	%\IEEEauthorblockN{5\textsuperscript{th} Fangfang Zhang}
	%\IEEEauthorblockA{\textit{Department of Electrical Engineering and Automation} \\
	%\textit{Qilu University of Technology}\\
	%Jinan, China \\
	%zhff4u@qlu.edu.cn}
	%}
	
	\maketitle
	
	\begin{abstract}
		All-gather collective communication is one of the most important communication primitives in parallel and distributed computation, which plays an essential role in many HPC applications such as distributed Deep Learning (DL) with model and hybrid parallelisms.
		To solve the communication bottleneck of All-gather, optical interconnection network can provide unprecedented high bandwidth and reliability for data transfer among the distributed nodes.
		However, most traditional All-gather algorithms are designed for electrical interconnection, which cannot fit well for optical interconnect systems, resulting in poor performance. 
		%All-to-all routing algorithms in optical interconnect systems focus on reducing the number of wavelengths required without considering how to reduce time. %However, most existing all-gather communication algorithms are designed for traditional electrical interconnect systems, which do not fit well for optical interconnect, thus resulting in poor performance. 
		This paper proposes an efficient scheme, called OpTree, for All-gather operation on optical interconnect systems. OpTree derives an optimal $m$-ary tree corresponding to the optimal number of communication stages, which achieves the minimum communication time.
		%by giving the number of nodes and wavelengths in optical interconnection systems. 
		We further analyze and compare the communication steps of OpTree with existing All-gather algorithms. Theoretical results exhibit that OpTree requires much less number of communication steps than existing All-gather algorithms on optical interconnect systems.  %The constraint on the optical power by OpTree is also discussed.
		%The routing tree is constructed using the number of grouping nodes $N^{\frac{1}{k^*}}$ and $k^*$ is the depth of the routing tree represented by optimal number of stages. 
		%The optimal stage $k^*$ and the number of grouping node $N^{k^*}$ decides optimal routing solution. We further derive the optimal communication steps and communication time of LBT by using the optimal stage $k^*$. 
		%Extensive simulations are conducted to evaluate OpTree. The experimental results show the effectiveness of our proposed OpTree. %and compared with one-stage model, OpTree can reduce communication time by 96.85\% on average.
		Simulation results show that OpTree can reduce communication time by 72.21\%, 94.30\%, and 88.58\% respectively compared with three existing All-gather schemes W\textsc{rht}, Ring, and NE, respectively.

	\end{abstract}
	%Optical interconnect is a good alternative replacement for electrical interconnect because of Wavelength Division Multiplexing (WDM) and high bandwidth.

	\section{Introduction}
	With the increasing number of GPUs integrated into systems, performing efficient communication among GPUs is crucial for running HPC applications. All-to-all communications (e.g., MPI All-gather operation) are heavily used to perform Fast Fourier Transforms (FFT) in many legacy scientific applications when data is distributed into multiple processes~\cite{khorassani2021adaptive}. All-gather operation is also gaining attention recently for performing model or hybrid parallelisms for training distributed Deep Neural Networks (DNNs) on GPU clusters~\cite{ben2019demystifying}. With the current software and hardware development for machine learning, optimizing All-gather communication on the current and next-generation dense-GPU systems is of great importance.
	
	Although many algorithms for All-gather operation have been proposed to improve the performance on electrical interconnect systems, there are still communication bottlenecks in current electrical interconnect systems~\cite{khani2021sip}.
	For example, in distributed model-parallel DNN training, frequent communications with large data transfer among distributed nodes are required. This may cause communication bottlenecks in electrical interconnect systems.
	With the development of silicon photonics, emerging optical networks can provide concurrent communication capacities due to Wavelength Division Multiplexing (WDM). They promise data transmission rates several orders of magnitude higher than current electrical networks. 
	With optical networks, the communication bottleneck can be avoided for many time-critical applications in datacenters, such as scientific visualization, high-speed simulation of climate change, parallel and distributed training of deep learning, etc.
	
	%With the recent development of CMOS-compatible optical devices~\cite{b4}, optical intra/inter-chip network connection is a promising alternative for electrical interconnect.  Optical interconnect can transmit data through a waveguide using different wavelengths enabled by leveraging Wavelength Division Multiplexing (WDM), thus providing high bandwidth, low transmission delay, and low power cost. Thus, we can solve the communication bottleneck of distributed DNN training on electrical interconnect system by replacing the electrical interconnect.
	%The purpose of the all-gather operation (all-to-all broadcast) is that data contributed by each worker is gathered onto every worker.
	%Hence, it is crucial to design scalable and efficient All-to-all algorithms to reduce the communication time in distributed DNN training.
	%We have investigated all-reduce operation on optical interconnect systems in our prior work W\textsc{rht}~\cite{wrht}. 
	%This paper focuses on two types of all-to-all collective communications: non-personalized all-to-all broadcast (All-gather) and personalized all-to-all scatter (all-to-all).  The research on traditional all-communicationto-all collective communications in optical interconnect systems is not designed for distributed DNN training. 
	
	Most traditional All-gather algorithms cannot fit well for optical interconnect systems because they do not take advantage of optical transmissions such as WDM.  Although there are some works on all-to-all broadcast in optical networks, most of them focus on fault tolerance~\cite{zhu2011reliable} or reducing the usage of wavelengths~\cite{liang2006general} rather than reducing communication time. %~\cite{bermond1996efficient}
	%If the wavelength resources can be properly unitized, the performance of all-to-all collective communications can be further improved when deploying distributed model-parallel training of DNN on optical interconnect system.  
	% which fails to take advantage of the parallel communication features using WDM in the optical network resulting in more communication time.
	%Besides, the current all-to-all broadcast communication schemes in the optical network require a large number of wavelengths without considering the real application. It can be seen that there is a gap in improving the performance of all-reduce in the optical interconnect system.  Deploying distributed training of DNN on optical interconnect system, we can greatly reduce the number of steps and communication time of all-reduce operation if we can reuse the wavelength and use the proper routing algorithm. 
	
	Inspired by the uniform routing and multi-stage model~\cite{liang2006general} in optical networks, we propose an efficient scheme, OpTree, for All-gather operation on optical ring interconnect systems, aiming at minimizing the communication time.
	%In this paper, we propose an efficient all-reduce scheme named \emph{WRHT} for an optical interconnect system interconnected with a ring topology. Our objective is to reduce the amount of communication time required to complete an all-reduce operation. 
	We use the ring topology because it is more feasible to be deployed and existing prototypes such as TeraRack~\cite{khani2021sip} have already implemented it.   
	%The reason why we use ring topology is that this topology is easy to set up and existing prototypes such TeraRack have already implemented it. 
	%The key idea of OpTree is to realize the all-gather operations by groups in a load balance way interactively so that the wavelength resources can be shared within each group and between groups with a minimum number of communication steps. 
	%The key idea of \emph{WRHT} is to create groups based on the available wavelengths with the number of grouping nodes $(2w+1)$ to increase the wavelength reuse so that the number of communication steps to realize all-reduce operation can be largely reduced. To the best of our knowledge, W\textsc{rht} is the first scheme for optimizing the all-reduce operation on optical interconnect systems. 
	To the best of our knowledge, OpTree is the first scheme to optimize the communication time for All-gather operation in optical interconnect systems.                  
	%, which can discover the relationships among communication time, number of nodes and wavelengths.
	
	Our main contributions are summarized as follows:
	\begin{itemize}
		\item We propose an efficient scheme OpTree based on $m$-ary tree structure for All-gather operation in optical interconnect networks. OpTree derives the optimal $m$-ary tree corresponding to the optimal number of communication stages and achieves the minimum communication time among all options of the $m$-ary tree.
		\item We analyze and compare the communication steps of OpTree with existing All-gather algorithms. Theoretical results show that OpTree requires a much smaller number of communication steps than existing All-gather algorithms on optical interconnect systems.  
		%The constraint of optical power in optical interconnect systems is also discussed.     
		%We derive the minimum number of communication steps and the optimal communication time to realize the all-gather on optical ring system. Compared with three existing all-gather approaches in electrical interconnect system, the number of communication steps is significantly reduced. 
		%We propose W\textsc{rht} scheme for distributed data-parallel DNN training in optical interconnect systems, which can obtain the minimal number of communication steps and communication time of all-reduce operation.  
		\item We evaluate OpTree with extensive simulations. Compared to existing All-gather algorithms W\textsc{rht}, Ring, and NE, OpTree can reduce communication time by 56.36\%, 92.76\%, and 85.54\% on average in optical interconnect systems with different number of nodes. OpTree can also reduce communication time by 88.06\%, 95.84\%, and 91.69\%, respectively, in a 1024-node optical interconnect system using different wavelengths. 
		%OpTree can reduce communication time by 72.21\%, 94.30\%, and 88.58\% on average in the simulated optical interconnect system compared with four existing All-gather schemes W\textsc{rht}, Ring, and NE, respectively. 
		%\item We further conduct scaling test and wavelength test for these All-gather algorithms in the optical interconnect system. Results exhibit the performance of OpTree is overall better than the other three All-gather algorithms and the performance of OpTree can be significantly improved by increasing the number of wavelengths. %particularly when the number of nodes is between 512 and 1024.      
		%Firstly, we compare W\textsc{rht} with three traditional all-reduce algorithms in optical interconnect system.  Results show that the communication time of W\textsc{rht} is reduced by 75.59\%,  49.25\%, and 70.1\% respectively. 
		%We also compare the performance of all-gather algorithms on the optical interconnect system with the electrical counterpart, which shows OpTree can reduce the communication time for all-gather operation by 86.69\% and 84.71\%.	%WRHT can reduce the all-reduce communication time by 97\% on average.
		%Secondly, we compare the all-reduce algorithms’ performance of the optical interconnect system with the electrical counterpart, which shows the communication time of W\textsc{rht} is reduced by 97.16\% and 96.74\% on average compared with two all-reduce algorithms in electrical interconnect system.
	\end{itemize}
	The rest of this paper is organized as follows. 
	Section~\ref{sec:related} presents the related works.
	Section~\ref{sec:LBFT} illustrates the optical interconnect architecture, motivation, the design of OpTree, the analysis of communication steps and communication time. %, constraint of optical power. 
	%Section~\ref{sec:performance} first illustrates the optical interconnect architecture, and motivation of our work. Then, the design of W\textsc{rht} scheme, communication steps, and communication time of W\textsc{rht} are demonstrated respectively.  %minimize the all-reduce communication time of distributed DNN training in the optical interconnect system. 
	Section~\ref{sec:evaluation} evaluates OpTree under different scenarios. 
	%Section~\ref{sec:limitation} discuss the limitation of this paper.
	Finally, Section~\ref{sec:conclusion} concludes the article.

	\section{RELATED WORK} \label{sec:related} 
	%Existing related works can be classified into two categories: performance optimization on all-gather algorithms on electrical interconnect system and all-to-all routing on optical interconnect system.
	Since the standardization of MPI, All-gather operation in electrical interconnect systems has been extensively studied by researchers and engineers. 
	%Various All-gather algorithms have been proposed to optimize the performance in different situations. 
	To reduce the bandwidth bottleneck of the electrical link, hierarchical algorithms~\cite{graham2011cheetah,traff2006efficient}, %, karonis2000exploiting
	multi-leader approach~\cite{kandalla2009designing}, and multi-lane communication~\cite{traff2020decomposing} were proposed. To minimize the number of links traversed during inter-node communication, a topology-aware collective algorithm~\cite{mirsadeghi2016topology} was proposed. %sack2012faster, kumar2004scaling
	A general optimization method that can be applied in different architectures was also proposed in ~\cite{faraj2006star}. %faraj2005automatic,thakur2005optimization
	However, they are not suitable for optical interconnect system.
	
	%We first analyze the related work on all-to-all Routing on optical interconnect system.
	Collective communication in optical networks was first proposed in~\cite{bermond1996efficient}. The authors discussed the bounds on the number of wavelengths and communication steps needed for broadcast and gossiping based on one-stage and multi-stage models. Many studies focused on specific topologies for one-stage collective communication to reduce the number of wavelengths, such as ring, torus, mesh, hypercubes, and trees~\cite{ sabrigiriraj2008all}. %beauquier1999all, narayanan1999all, kumar2004scaling,  zhang1999scheduling
	In~\cite{pascu2009conflict}, the authors proposed a one-stage broadcasting model in WDM networks considering the tap-and-continue feature of optical nodes. Many follow-up investigations were carried out based on multi-stage models in different topologies. In~\cite{opatrny2003uniform}, the authors studied the uniform all-to-all routing problem using the multi-stage model in a symmetric directed ring to reduce the number of wavelengths. %Furthermore, many researches were conducted based on multi-stage model in different topologies. 
	%In~\cite{gu2003multihop}, the authors addressed the collective communication problem in multi-stage WDM networks for some special topologies: lines, ring, 2-D square tori, and 3-D square tori. 
	In~\cite{liang2006general}, the authors proposed a general multi-stage routing and wavelength assignment to reduce the number of wavelengths in the optical network. 
	%Most research on collective communication in optical networks are focusing on reducing the numbers of wavelengths by proposing multi-stage models.
	Our work differs from these works as we aim to optimize the time for All-gather operation in optical interconnect systems. 
	%As far as we know, there is first study exploring optimizing the time of All-gather operation in optical interconnect systems, where this study is bridging the gap.

	%Nowadays, it is still in its early stage to use optical networks as the interconnect system for distributed deep learning.
	%To date, the use of optical networks in distributed deep learning systems is still in its early stages. 
	%To the best of our knowledge, W\textsc{rht} is the first scheme proposed for optimizing the all-reduce operations on optical interconnect systems. %Other related works include the study of all-to-all broadcast communication in WDM optical network~\cite{b18}. Most existing solutions formulate the routing and wavelength assignment problems as integer linear programming optimization problems and focus on reducing the usage of wavelengths without considering the real workload.

	\section{The OpTree Scheme}  \label{sec:LBFT} 
	\subsection{Optical Interconnect Architecture} \label{sec:architecture} 
\begin{figure}[hpbt]
		\centering
		\vspace{-4mm}
		\includegraphics[scale=0.35]{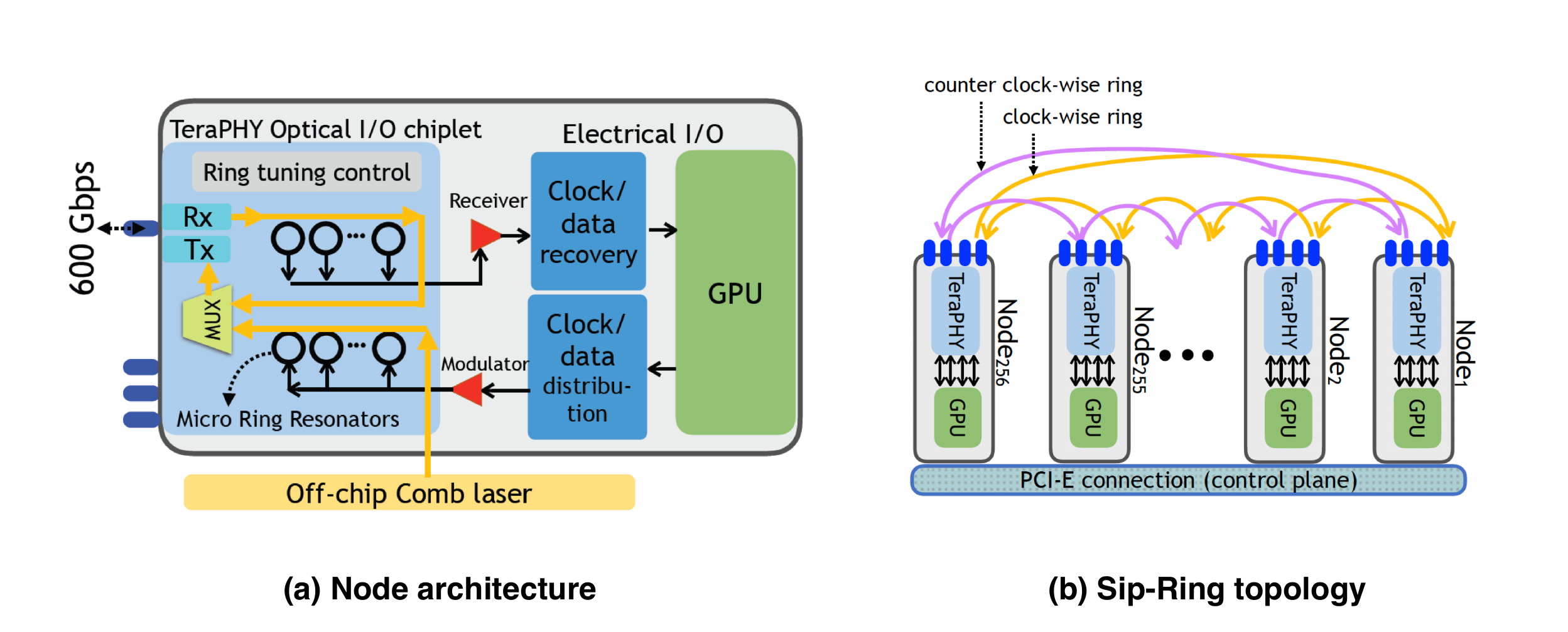}
		{
			\vspace{-5mm}
			\caption{Optical interconnection architecture: (a) TeraRack node, (b) Double ring topology~\cite{khani2020terarack}.}
			\label{fig:terarack}
		}
		\vspace{-2mm}
	\end{figure}

	We assume OpTree adopts the microring resonators (MRRs)-based optical switch called TeraRack~\cite{khani2020terarack}, while it is applicable to other similar optical interconnect systems.
	Fig.~\ref{fig:terarack}  (a) shows a TeraRack node and its components based on TeraPHY silicon photonics technology. We assume homogeneous GPUs are used as the computing devices. 
	There are four optical interfaces on a TeraRack node. Each interface has 64 micro-ring resonators to select and forward any subset of 64 wavelengths.
	The laser is based on a comb laser source called SuperNova Light Supply~\cite{khani2020terarack}.
	On the transmit side (Tx), an off-chip comb laser generates light that is steered into the node via a fiber coupler toward an array of MRRs modulating the accelerator’s transmitting data at 40 Gbps per wavelength. 
	On the receiver side (Rx), the second array of MRRs selects the wavelengths targeted at the accelerator and passes through the remaining wavelengths. 
	Fig.~\ref{fig:terarack} (b) shows that the nodes interconnect with their adjacent nodes in a ring topology (two rings for two directions are shown for clarity).
	In the data plane, the traffic is transmitted across four single-mode fiber rings: two clockwise and two counterclockwise. The Routing and Wavelength Assignment (RWA) configuration is done in the control plane, where the wavelengths can be dynamically placed around the fiber rings. 
	%This unique characteristic simplifies the control-plane logic with which datacenter optical designs have grappled for years. 
	The details of the system parameters are shown in Section IV.
	
	\subsection{Preliminary}\label{sec:preliminary}  
	The role of the All-gather operation is to make every worker concurrently broadcast the data item until every worker receives the data item from each other. In an optical interconnect system, the All-gather operation corresponds to all-to-all broadcast routing using concurrent optical communications by multiple wavelengths. %One straightforward approach for all-to-all routing in optical network is one-stage model~\cite{bermond1996efficient}, by which the data item from each worker is transmitted to all other workers directly in the form of light  without any intermediate relay nodes. However, 
	When the network size is large, the basic approach of one-stage model can lead to an unrealistically large number of wavelengths when realizing all-to-all routing, as can be seen in the following lemma~\cite{liang2006general}. 
	\begin{lemma}\label{lm:w}
		In an $N$-node optical interconnect system, the minimum number of wavelengths needed for all-to-all routing under one-stage model is $\lfloor\frac{N^2}{4}\rfloor$ in line and $\lceil\frac{N^2}{8}\rceil$ in ring. 
	\end{lemma}
	%requires  $\lfloor\frac{N^2}{4}\rfloor$ wavelengths in the optical line network and requires $\lceil\frac{N^2}{8}\rceil$ wavelengths in the optical ring network using one-stage mode. 
	Since the number of wavelengths in the optical interconnect system is limited, multi-stage ($k$-stage, $k \geq 2$) model can be used to reduce the number of wavelengths, which takes $k$ communication stages to complete all-to-all optical communication~\cite{bermond1996efficient}. For the $k$-stage model, the optical signals must be converted to electronic form $k-1$ times before the All-gather operation is completed. 
	However, most designs for multi-stage model only focus on reducing the number of wavelengths but fail to address the performance of All-gather. 
	Therefore, we propose an efficient scheme, OpTree, for All-gather operation on optical interconnect systems based on $m$-ary tree~\cite{tolentino2022twin} structure. OpTree can optimize the communication time of All-gather.
	In graph theory, an $m$-ary tree is a tree structure in which each internal node has no more than $m$ children. A binary tree is a special case where $m = 2$. 
	\begin{lemma}\label{lm:l}
		For an $m$-ary tree with height $k$, the upper bound for the maximum number of leaves is $m^k$.
	\end{lemma}

	\begin{figure}[hpbt]
		\vspace{-2mm}
		\centering
		\includegraphics[scale=0.42]{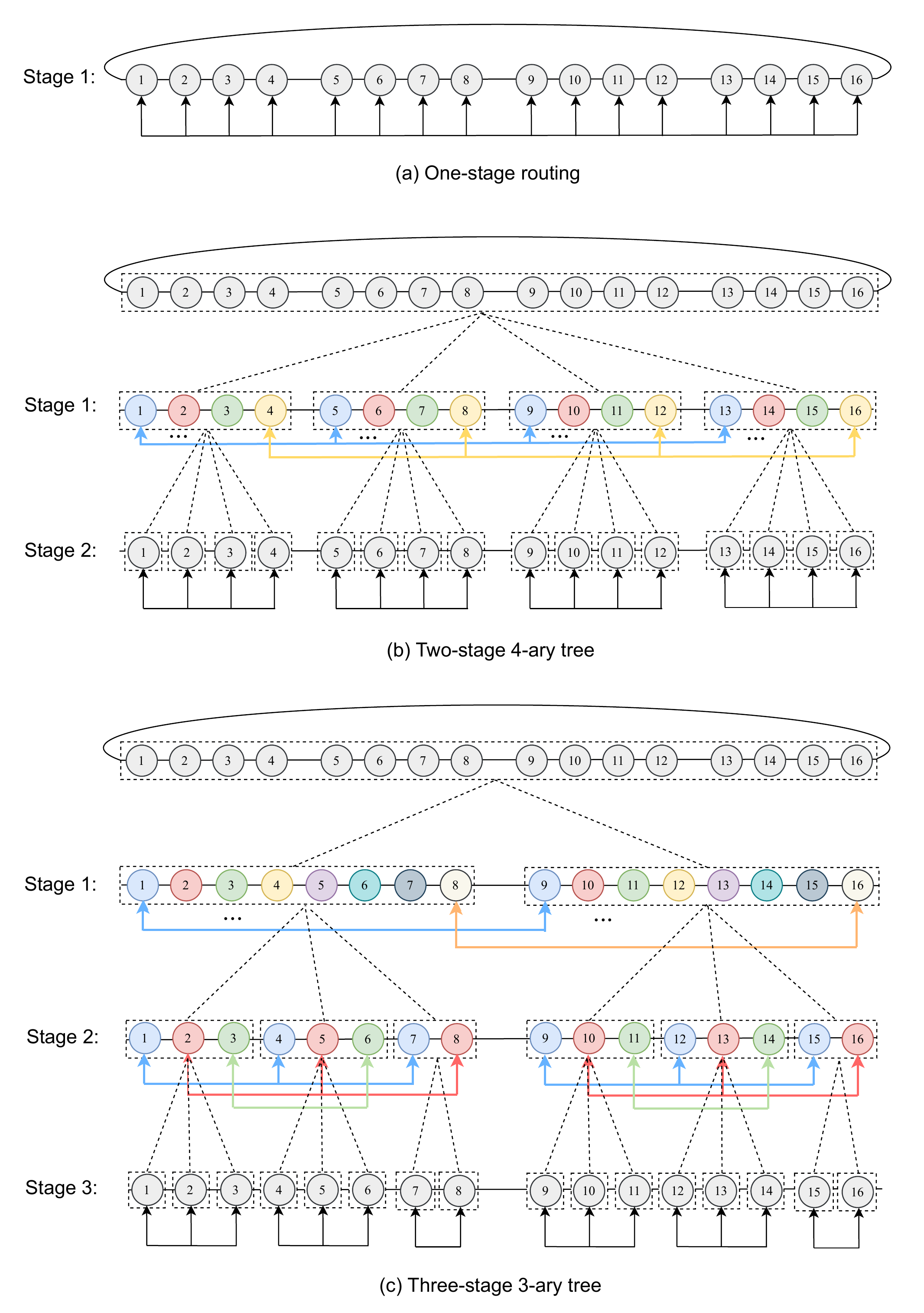}
		{
			\vspace{-8mm}
			\caption{(a) One-stage, (b) Two-Stage 4-ary tree, and (c) Three-stage 3-ary tree.}
			\label{fig:LBFT_}
		}
		\vspace{-3mm}
	\end{figure}

	\subsection{Motivation}
	We use a motivation example to illustrate the trade-off between the communication time and the number of stages for our design. We assume that the system has 16 nodes based on the architecture of Fig.~\ref{fig:terarack}, the number of available wavelengths is 2, and every node has an amount of data $d$ to collect from all the other nodes. We compare the communication time for the following three schemes: (a) One-stage routing, (b) Two-stage 4-ary tree, and (c) Three-stage 3-ary tree.
	%For a given number of nodes and available wavelengths in the optical interconnect system, we can have different routing solutions to finish the all-gather operation. %The research question is how to find the load balancing routing that can have the minimum communication time.  
	%We use one stage routing and two different $m$-ary trees of OpTree to illustrate how we find the optimal $m$-ary tree that has minimum communication steps and communication time in OpTree.  
	%%to explain the different communication steps of all-gather operation is .
	%In the motivation example, we assume the system has 16 nodes based on the architecture of Fig.~\ref{fig:terarack}, and the available number of wavelengths is 2. Every node has an amount of transferred data of $d$ to collect from all the other nodes. We use the term level of the tree and the stage of OpTree interchangeably.
	
	{\bf One-stage routing}: As shown in Fig.~\ref{fig:LBFT_} (a), one stage routing conducts all-to-all broadcast among all nodes without any intermediate relay nodes. This requires $\lceil\frac{16^2}{8}\rceil = 32$ wavelengths according to Lemma~\ref{lm:w}. Since the number of available wavelengths is 2, it takes $\lceil\frac{32}{2}\rceil = 16$ communication steps (time slots) to finish the All-gather operation. 
	
	{\bf Two-stage 4-ary tree}: As shown in Fig.~\ref{fig:LBFT_} (b), the whole group of 16 network nodes is taken as a super root node of a full 4-ary tree, which is partitioned into 4 children (subgroup) in stage 1. Each child tree node contains 4 network nodes. 
	During stage 1, the network nodes on the $i$th position in each sibling node perform all-to-all routing using one stage model along the ring, and the amount of data sent is $d$. 
	So, the network nodes 1, 5, 9, and 13 (marked in blue) perform all-to-all routing as a subset, and 2, 6, 10 and 14 (marked in red) perform all-to-all routing as a subset, and so on. Since these four subsets share the optical links in the ring, the wavelength requirement for stage 1 is $4\cdot\lceil\frac{4^2}{8}\rceil = 8$. This requires 4 communication steps (time slots) to complete. In stage 2, each subgroup (child node of the root node) is further partitioned into 4 children, with each child node having one network node (leaf of the 4-ary tree). During stage 2, sibling nodes conduct all-to-all routing using one stage model on the segment of the ring, and the amount of data sent is $4d$.  So, network nodes 1, 2, 3, and 4 perform all-to-all routing as a subset, and 5, 6, 7 and 8 perform all-to-all routing as a subset, and so on. To achieve load balancing in each stage, each wavelength is loaded with a data item of size $d$. Since these four subsets do not share the optical links in the segment of the ring (line), the wavelength requirement of stage 2 is $4\cdot\lfloor\frac{4^2}{4}\rfloor=16$. This requires 8 communication steps (time slots) to finish. Therefore, the total number of communication steps by two-stage 4-ary tree is $\lceil \frac{8}{2} \rceil+\lceil \frac{16}{2} \rceil= 12$.
	
	{\bf Three-stage 3-ary tree}:  
	As shown in Fig.~\ref{fig:LBFT_} (c), the three-stage 3-ary tree scheme allows each node to contain at most 3 children with depth of $log_3 16=3$ (i.e., 3 stages). Accordingly, we can calculate the wavelength requirements for stages 1, 2, and 3 are $8\times\lceil\frac{2^2}{8}\rceil=8$, $2\times3\times\lfloor\frac{3^2}{4}\rfloor=12$, and $6\times\lfloor\frac{3^2}{4}\rfloor=12$ respectively. Therefore, the total number of communication steps to finish the All-gather operation by the three-stage 3-ary tree is $\lceil \frac{8}{2} \rceil+\lceil \frac{12}{2} \rceil+\lceil \frac{12}{2} \rceil=16$.
	
	We can see that 4-ary tree scheme has the least number of communication steps among these three schemes. One-stage model only uses one stage but requires more time slots than 4-ary tree scheme, whereas 3-ary tree scheme uses more stages than 4-ary tree scheme, but still requires more time slots than 4-ary tree scheme. So, there is a trade-off between communication time and the number of stages by $m$-ary tree to conduct All-gather operation on optical interconnect systems. The challenging problem is: for a given number of nodes and available wavelengths in the optical interconnect system, how to find the optimal $m$-ary tree with the optimal number of stages that minimizes communication steps and communication time among all the options of the $m$-ary tree. 

	\subsection{Design of OpTree}
	To address the above challenge, we design an efficient scheme OpTree, represented by $m$-ary tree structure with the number of All-gather stages represented by the tree depth of $k$. We first present the routing algorithm and then analyze the communication steps and communication time by OpTree.
	\subsubsection{Routing of OpTree}
\begin{figure}[!hpbt]
		\vspace{-3mm}
		\includegraphics[scale=0.41]{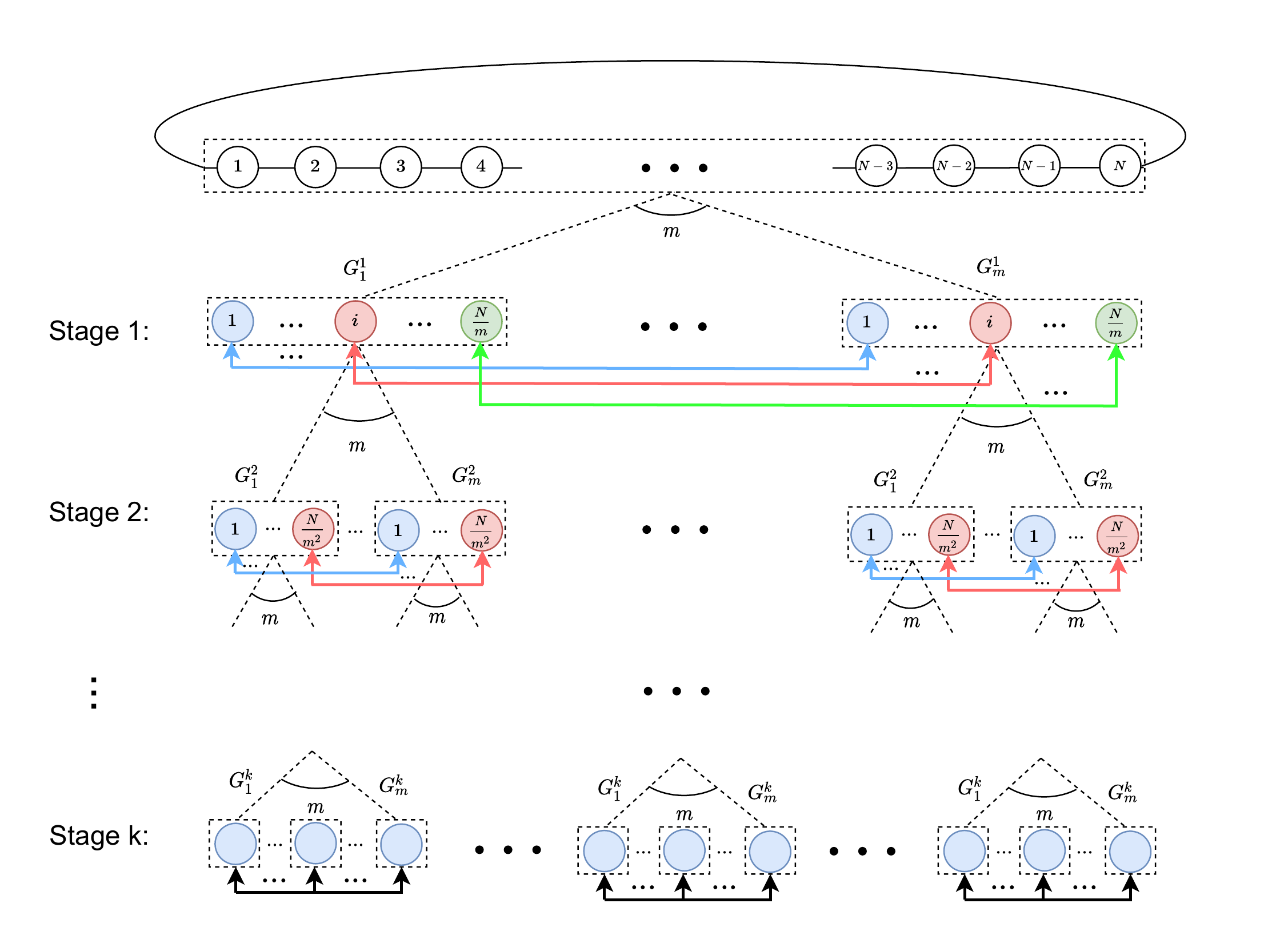}
		{
			\vspace{-7mm}
			\caption{The working principle in OpTree algorithm.}
			\label{fig:LBFT}
		}
		\vspace{-1mm}
	\end{figure}

	Assume that the number of nodes in the optical interconnect system is $N$, and the number of available wavelengths is $w$. 
	Therefore, the $i$th level of the $m$-ary tree corresponds to the $i$th stage of All-gather operation. As illustrated in Fig.~\ref{fig:LBFT}, the $k$-level $m$-ary tree is constructed by recursively partitioning the group of network nodes into $m$ sub-groups as child tree nodes. According to Lemma 2, each parent tree node has at most $m = {N^{\frac{1}{k}}}$ child tree nodes, with each child tree node containing $N^{\frac{k-1}{k}}$ network nodes at the $i$th level of $m$-ary tree. The depth of the $m$-ary tree is $k=log_mN$, which is also the number of stages for All-gather operation.
	The detailed working principle of OpTree is described below. 

	\textbf{Stage $1$:} 
	Initially, the whole group of $N$ network nodes is regarded as the root tree node, which is divided into $m$ child tree nodes denoted as $G_1^1, ..., G_m^1$ with each child node containing $\lceil \frac{N}{m} \rceil$ network nodes. During stage $1$, the network node on the $i$th ($i \in [1,\lceil \frac{N}{m} \rceil]$) position in each of these $m$ sibling nodes can form a subset to perform all-to-all routing by one-stage model. For example, the blue node (node 1) in each of $G_1^1, ..., G_m^1$ broadcasts its data item with size $d$ to all the other blue nodes by one-stage model along the ring. %To differentiate the logical and physical groups, we first use dotted rectangles to denote the physical groups. Among the physical groups, the nodes with the same color belong to the nodes in the logical group, as seen in Fig.~\ref{fig:LBFT}.

	\textbf{Stage $j$:}
	Recursively, each node in stage $j-1$ containing $\lceil \frac{N}{m^{j-1}}\rceil$ network nodes is further partitioned into $m$ child nodes denoted as $G_1^j, ..., G_m^j$ with each child node having $\lceil \frac{N}{m^j}\rceil$ network nodes, until each child node only contains one network node at the last stage $k$.  %, i.e. $\lceil \frac{N}{m^k}\rceil=1$
	During stage $j$, the network node in the $i$th ($i \in [1,\lceil \frac{N}{m^j} \rceil]$) position in each of its $m$ sibling nodes can form a subset to perform all-to-all routing by one-stage model. For example, during stage 2, the blue node (node 1) in each of $G_1^2, ..., G_m^2$ broadcasts its data item to all the other blue nodes among sibling nodes. During stage $k$, each leaf node contains only one network node, so $G_1^k, ..., G_m^k$ exchange data with each other among sibling nodes by one-stage model. 
	%form new logical groups with nodes at the same positions within the new physical groups, . They need to exchange the amount of transferred data $N^{\frac{j-1}{k}}d$ with other nodes in the logical group by one-stage model of the line using $\lfloor \frac{(N^{\frac{1}{k}})^2}{4}\rfloor$ wavelengths. 
	In stage $j$, the number of network nodes in each  tree node is $\lceil \frac{N}{m^j} \rceil$, and each network node has $m^{j-1} d$ data to send by $m^{j-1}$ wavelengths for load balance. %So, the required number of wavelengths on stage $j$ is $m^{j-1}\times \lceil \frac{N}{m^j} \rceil \times \lfloor \frac{m^2}{4} \rfloor =\lfloor \frac{m^2}{4} \rfloor \times \lceil \frac{N}{m} \rceil$, and the communication steps on stage $j$ is $\lfloor \frac{m^2}{4w} \rfloor \times \lceil \frac{N}{m} \rceil$. 

	\subsubsection{Analysis of Communication Steps}
	In order to achieve load balancing, OpTree algorithm specifies that each wavelength carries the same amount of data $d$ during each communication step. Therefore, the communication time of All-gather by OpTree is mainly determined by the toal number of communication steps, which we calculate as follows.
	%and then derive the optimal $m$-ary tree corresponding to the optimal number of communication stages achieving the minimum communication time among all the $m$-ary tree options.  
	\begin{theorem}\label{th:cs}
		In an $N$-node optical ring interconnect system with $w$ available wavelengths, the total number of communication steps needed to perform All-gather operation during $k$ stages ($k \geq 2$) by OpTree is $  \lceil \frac{(2k-1)N^{1+\frac{1}{k}}}{8w}  \rceil$. 
	\end{theorem}
	\begin{proof}
		%The all-to-all routing for a line and a ring under one hop model has been studied and the minimum number of wavelengths needed is $\lfloor \frac{N^2}{4} \rfloor$~\cite{Efficient}, $\lceil \frac{N^2}{8} \rceil$~\cite{ring}, respectively. 
		%We first prove the theorem is correct when k = 2 in LBFT. When k = 2, the number of wavelengths needed for k stages of LBFT is $ \frac{3N^{\frac{3}{2}}}{8}$. On Stage 1, N nodes are divided into $N^{\frac{1}{2}}$ groups with each group having $N^{\frac{1}{2}}$ nodes. Stage 1 needs $\lceil \frac{(N^{\frac{1}{2}})^2}{8} \rceil N^{\frac{1}{2}}$ wavelengths, and Stage 2 needs $\lfloor \frac{(N^{\frac{1}{2}})^2}{4} \rfloor N^{\frac{1}{2}}$ wavelengths. Adding the number of wavelengths of two stages together, we can obtain the required number of wavelengths for two stage of LBFT $ \frac{3N^{\frac{3}{2}}}{8}$. So, the theorem holds when k = 2. 
		In stage 1, since there are $\lceil \frac{N}{m} \rceil$ subsets sharing the optical links, the number of communication steps is $ \lceil\frac{m^2}{8w}\rceil \times \lceil \frac{N}{m} \rceil$. For each of the subsequent $k-1$ stages, the required number of communication steps in stage $j$ ($j \in [2, k]$) is $m^{j-1}\times \lceil \frac{N}{m^j} \rceil \times \lfloor \frac{m^2}{4w} \rfloor =\lfloor \frac{m^2}{4w} \rfloor \times \lceil \frac{N}{m} \rceil$. 
		%we use $j$, $j \in [1, k-1] $ to denote the first $k-1$ stages. The number of wavelengths On Stage $j$ is $\lceil \frac{(N^{\frac{1}{k}})^2}{8} \rceil N^{\frac{j}{k}}$. On Stage $k$ of OpTree, the number of wavelengths is $\lfloor \frac{(N^{\frac{1}{k}})^2}{4} \rfloor N^{\frac{1}{k}}$.  
		Let $S$ be the total number of communication steps for all the $k$ stages of OpTree, which can be obtained as follows.  
		\begin{equation}\label{eq:ns}
			\begin{split}
				S= \lceil\frac{(2k-1)N^{1+\frac{1}{k}}}{8w}\rceil. 
			\end{split}
		\end{equation}
		%As $m = {N^{\frac{1}{k}}}$, the total number of communication steps during all the $k$ stages of OpTree are $S=\lceil \frac{(2k-1)N^{\frac{k+1}{k}}}{8w}  \rceil$. 
		Therefore, the theorem holds.
	\end{proof}
	By Theorem~\ref{th:cs}, the optimal $m$-ary tree corresponding to the optimal number of communication stages can be derived as follows, which can achieve the minimum communication time among all the $m$-ary tree options.

	\begin{theorem}\label{eq:stages}	
		In an $N$-node optical ring interconnect system with $w$ available wavelengths, the optimal number of communication steps by OpTree for finishing the All-gather operation is obtained at $k^*$ =  $ \left[ \frac{ln N + \sqrt{ln N (ln N-2)  }  }{2} \right]$.
		%and the optimal number of communication steps by OpTree is $S^*= \lceil  \frac{(2k^*-1)N^{1+\frac{1}{k^*}}}{8w}  \rceil$.
	\end{theorem}
	
	\begin{proof}
		Assuming that $S$ is a continuous function and $k$ is a variable in Eq.~(\ref{eq:ns}), the number of communication steps is minimized when $\frac{\partial S}{\partial k} = 0$. It can be derived that
		\begin{equation}\nonumber
			\frac{\partial S}{\partial k} = \frac{N^{1+\frac{1}{k}}}{4w} - \frac{N^{1+\frac{1}{k}}\cdot ln N\cdot (2k-1)}{8 w k^2}.
		\end{equation}
		Let $\frac{\partial S}{\partial k} = 0$. $S$ is minimized when 
		\begin{equation}
			k^* =  \left[ \frac{ln N + \sqrt{ln N (ln N -2)  } }{2} \right], 
		\end{equation}
		where $[\cdot]$ represents the integer rounding operation. %By replacing $k$ with $k^*$ in Eq.~(\ref{eq:ns}), the optimal number of communication steps by OpTree is $S^*= \lceil  \frac{(2k^*-1)N^{1+\frac{1}{k^*}}}{8w}  \rceil$.
		%Therefore,  the optimal number of communication steps by LBFT for finishing the all-gather and all-to-all operations is obtained at $ k = \frac{ln(N)\left(3 + \sqrt{9ln(N) + 12} \right)}{2}$. 
	\end{proof}
	%By Lemma~\ref{eq:stages}, the corresponding minimum communication time among all the OpTree options can be derived by replacing $k$ with $k^*$ in Eq.~(\ref{eq:ns}) as follows.  
	
	%\begin{theorem}\label{th:steps}
	%In an $N$-node optical ring interconnect system with $w$ available wavelengths, the optimal number of communication steps by OpTree is $ \lceil  \frac{(2k^*-1)N^{1+\frac{1}{k^*}}}{8w}  \rceil$. \end{theorem}
	%\begin{proof}
	%	According to the proof for Lemma~\ref{eq:stages}, the optimal number of communication steps by OpTree is minimized when $k$ = $k^*$. Hence, by replacing k with $k^*$ in Eq.~(\ref{eq:ns}), the optimal number of communication steps by OpTree is $ \lceil  \frac{(2k^*-1)N^{\frac{k^*+1}{k^*}}}{8w}  \rceil$.
	%\end{proof}
	%The communication time of all-gather operation is dominated by the number of communication steps since MRRs should be reconfigured before each communication step.
	
	Table~\ref{tab:comparison} summarizes the number of communication steps between different algorithms for the All-gather operation, including the traditional Ring All-gather~\cite{chen2005performance}, Neighbor Exchange (NE)~\cite{chen2005performance}, W\textsc{rht}\footnote{W\textsc{rht} algorithm is an algorithm designed for All-reduce~\cite{dai2022wrht}, which can be extended to the All-gather operation. $\theta = \lceil \log_{p}N \rceil$ and $p = 2w+1$.}~\cite{dai2022wrht}, and One-stage model.
	\begin{table}[!ht]
		\centering
		\vspace{-1mm}
		\caption{Communication step comparison for different All-gather algorithms in optical interconnect systems.}
		\label{tab:comparison}
		\begin{center}
			\begin{tabular}{ccc}
				\hline
				\multicolumn{1}{c}{\multirow{2}{*}{Algorithm}} & \multicolumn{1}{c}{\multirow{2}{*}{Communication steps}} & Number of steps \\ \cline{3-3} 
				\multicolumn{1}{c}{}                           & \multicolumn{1}{c}{}                                    & N = 1024, $w$ = 64       \\ \hline
				Ring                                              &        $N-1$                                                 &   1023    \\
				NE                                              &     $\frac{N}{2}$  &     512     \\
				
				W\textsc{rht}                                              &   $\left\lceil\frac{N-p}{p-1}\right\rceil+ \left\lceil \frac{(\theta-1)N}{p}\right\rceil+1$ &   259    \\
				One-Stage                                              &     $\lceil \frac{N^2}{8w} \rceil$  &     128     \\
				OpTree                                              &     $\lceil  \frac{(2k^*-1)N^{1+\frac{1}{k^*}}}{8w}  \rceil$  &     70 ($k^*$ = 7)     \\
				\hline
			\end{tabular}
		\end{center}
		\vspace{-2mm}
	\end{table}

	\subsubsection{Communication Time of OpTree}
	Since we obtain the optimal number of communication steps, we can further derive the total communication time to finish the All-gather operation by OpTree, denoted as $T_{comm}$:  
	%Since we have obtained the optimal number of communication steps for the all-gather operation, we can further derive the total amount of transferred data of OpTree and calculate the communication time to finish the all-gather. We denote $a$ as the delay of O/E/O conversion and reconfiguration delay of the MRRs, $B$ as the bandwidth per wavelength, and $d$ as the amount of transferred data to be received initially for each node. Thus the communication time of OpTree, denoted as $T_{comm}$, can be calculated as:   
	\begin{equation}\label{eq:wrht}
		\begin{split}
			T_{comm}&= (\frac{d}{B} + a)\lceil  \frac{m log_mN N}{4w}  \rceil,
		\end{split}
	\end{equation}%\frac{d S  }{B} + aS      (\frac{d}{B} + a)\lceil  \frac{m log_mN N}{4w}  \rceil
	%line_unbalance:  
	where $a$ is the O/E/O conversion delay and the reconfiguration delay of the MRRs, $B$ is the bandwidth per wavelength, $d$ is the amount of transferred data to be received initially for each node, and $S$ represents the total number of communication steps. Since the parameters of $d$, $B$, and $a$ are all constant values, the optimal communication time using OpTree can be achieved when the number of communication steps is minimized according to Theorem 2, as shown below.     
	%the communication step of \textit{WRHT} is denoted as $F(N, w, 2w+1)$, 
	
	\begin{theorem}\label{th:op_time}
		In an $N$-node optical ring interconnect system with $w$ available wavelengths, the optimal communication time by OpTree for All-gather operation is $(\frac{d}{B} + a)\lceil  \frac{N m^* log_{m^*}N }{4w}  \rceil$. 
	\end{theorem}%(\frac{d}{B} + a)\lceil  \frac{(2k^*-1)N^{1+\frac{1}{k^*}}}{8w}  \rceil

	\section{Evaluation}   \label{sec:evaluation} 
	%From the above two computation orders, we can have at least two parallel schemes during the convolution computation.   
	%In the input major order, AlexNet (21.622M), VGG16 (129.582M), GoogLeNet (5.269M) and ResNet50 (2.05M)
	\subsection{Simulation setup}
	We use the same optical interconnect simulator as~\cite{dai2022wrht} to test the performance of OpTree. 
	%Since the simulator is based on the all-reduce algorithms, we modify the simulator and further implements the proposed OpTree and three all-gather algorithms.
	The bidirectional ring topology for the optical interconnect system is the same as tera-Rack~\cite{khani2020terarack}. The default number of wavelengths is 64, and the bandwidth per wavelength is 40 Gbps. The optical transmission packet is 128 bytes, and the flit size is 32 bytes. The data type for the All-gather operation is set to be float32, and each node in the optical interconnect system has one GPU. 
	The reconfiguration delay of MRRs is 25 $\mu s$~\cite{khani2021sip}, and the conversion latency of O/E/O is one cycle/flit~\cite{dai2021performance}. 

	\subsection{Optimal $k$-stage $m$-ary Tree}
\begin{figure}[!ht]
		\vspace{-3mm}
		\centering
		\includegraphics[scale=0.5]{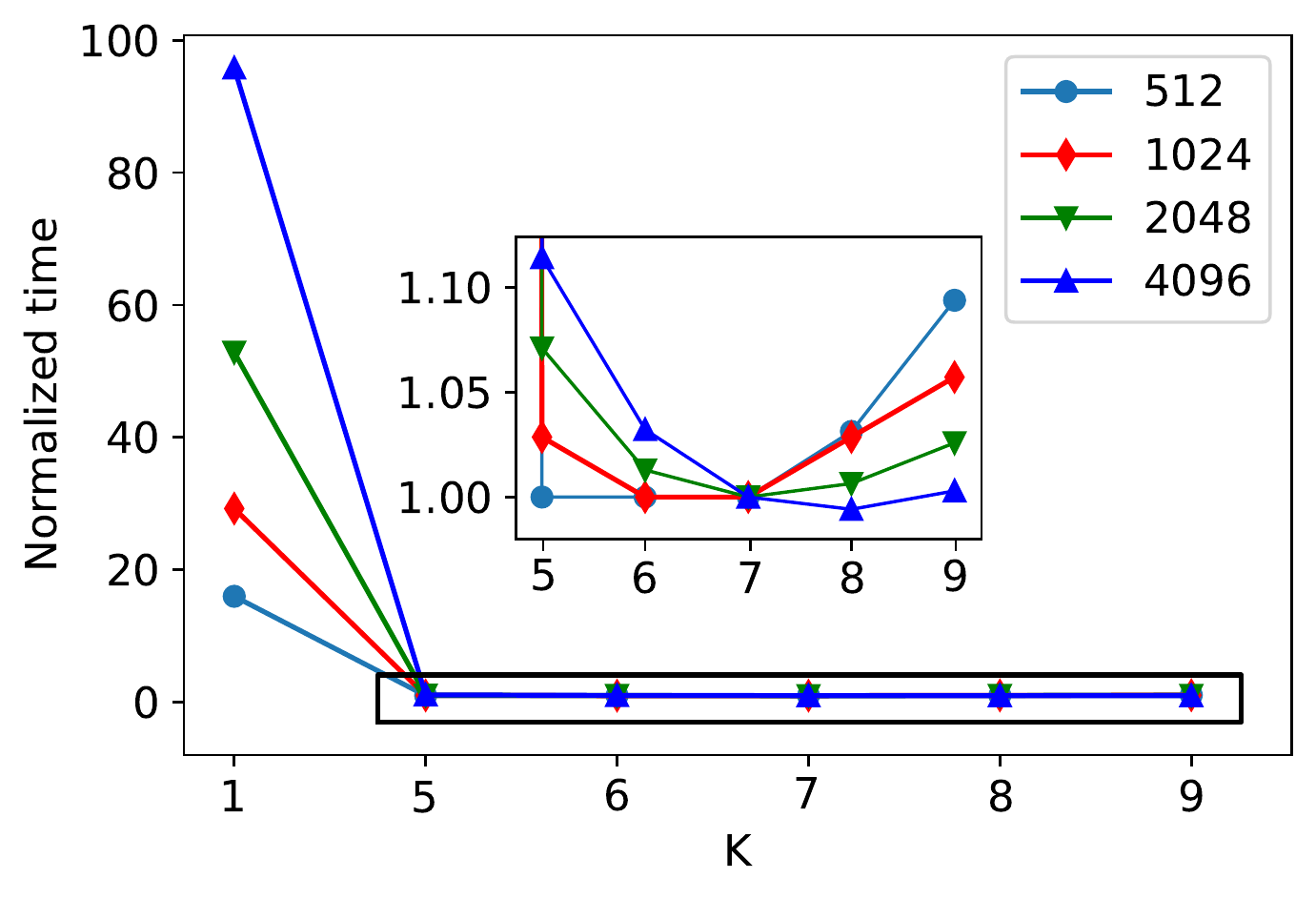}
		{
			\vspace{-3mm}
			\caption{Performance comparison of OpTree with different depths (denoted by $k$) for different number of nodes.}
			\label{fig:optree}
		}
	\end{figure}

	In this set of simulations, we verify that OpTree can find an optimal $m$-ary tree for All-gather operation with minimum communication time among different $m$-ary tree options under different numbers of nodes.  
	%In OpTree, the depth of the $m$-ary tree with $N$ nodes is $k$, where $k = log_mN$. Therefore, we can use $k$ to denote the different $m$-ary trees for different numbers of nodes.
	We vary the number of nodes from 512 to 4096 with the message size set to 4M. All results in Fig.~\ref{fig:optree} are normalized by dividing the optimal result of OpTree.
	
	Fig.~\ref{fig:optree} shows four line charts representing the performance comparison of $m$-ary trees with different depths (denoted by $k$) for All-gather operation with 512, 1024, 2048, and 4096 nodes. $k=1$ means we adopt the one-stage model for the All-gather operation. The inset of Fig.~\ref{fig:optree} is a magnified part of four line charts under different numbers of nodes. It can be seen that there is a trade-off between the depth of $m$-ary trees and the number of nodes in the optical interconnect system. With the increasing tree depth from 5 to 9, the performance for 512 nodes stays still at first and then increases dramatically, while the performance for 1024 nodes first decreases and stays the same, then increases rapidly afterwards. The performance for 2048 nodes keeps dropping to the bottom at depth 7 and then increases. The performance for 4096 nodes is similar to that of 2048 nodes, as it first decreases dramatically, reaches the bottom at depth 8, and then increases slightly. It can be seen that the optimal performance can be achieved at depths 6, 6, 7, and 8 for 512, 1024, 2048, and 4096 nodes, respectively. This is consistent with the theoretical results obtained from Theorem~\ref{eq:stages} and Theorem~\ref{th:op_time}, which shows the correctness of OpTree. 
	Compared with the one-stage model in a ring for different numbers of nodes, Optree can reduce communication time by 96.85\% on average. 

	\subsection{Performance Comparison}
\begin{figure}[!ht]
		\vspace{-2mm}
		\centering
		\includegraphics[scale=0.5]{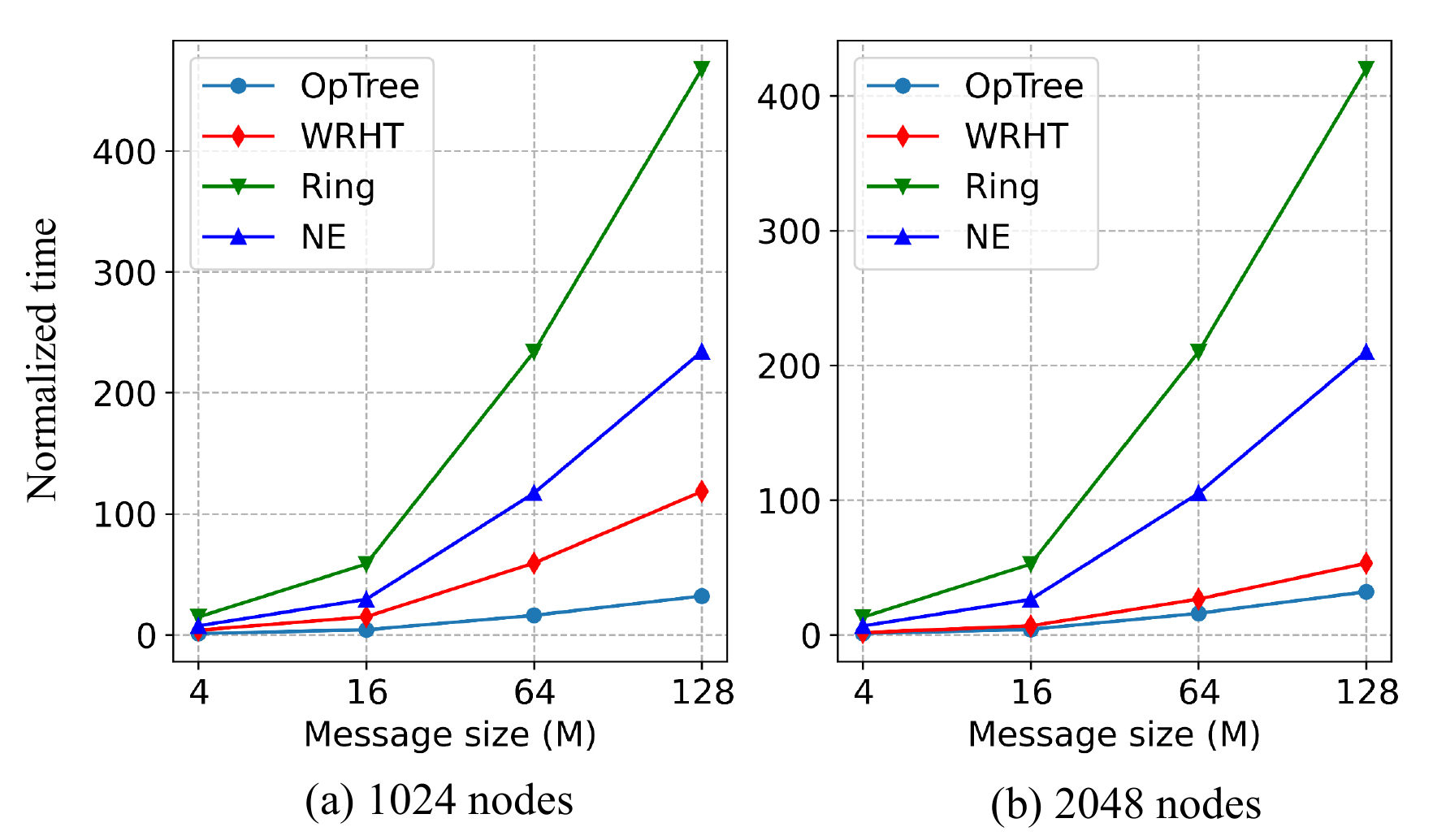}
		{
			\vspace{-5mm}
			\caption{Performance comparison of different All-gather algorithms in optical interconnect system with different numbers of nodes}
			\label{fig:nodes}
		}
		\vspace{-2mm}
		
	\end{figure}

\begin{figure}[!ht]
		\vspace{-6mm}
		\centering
		\includegraphics[scale=0.5]{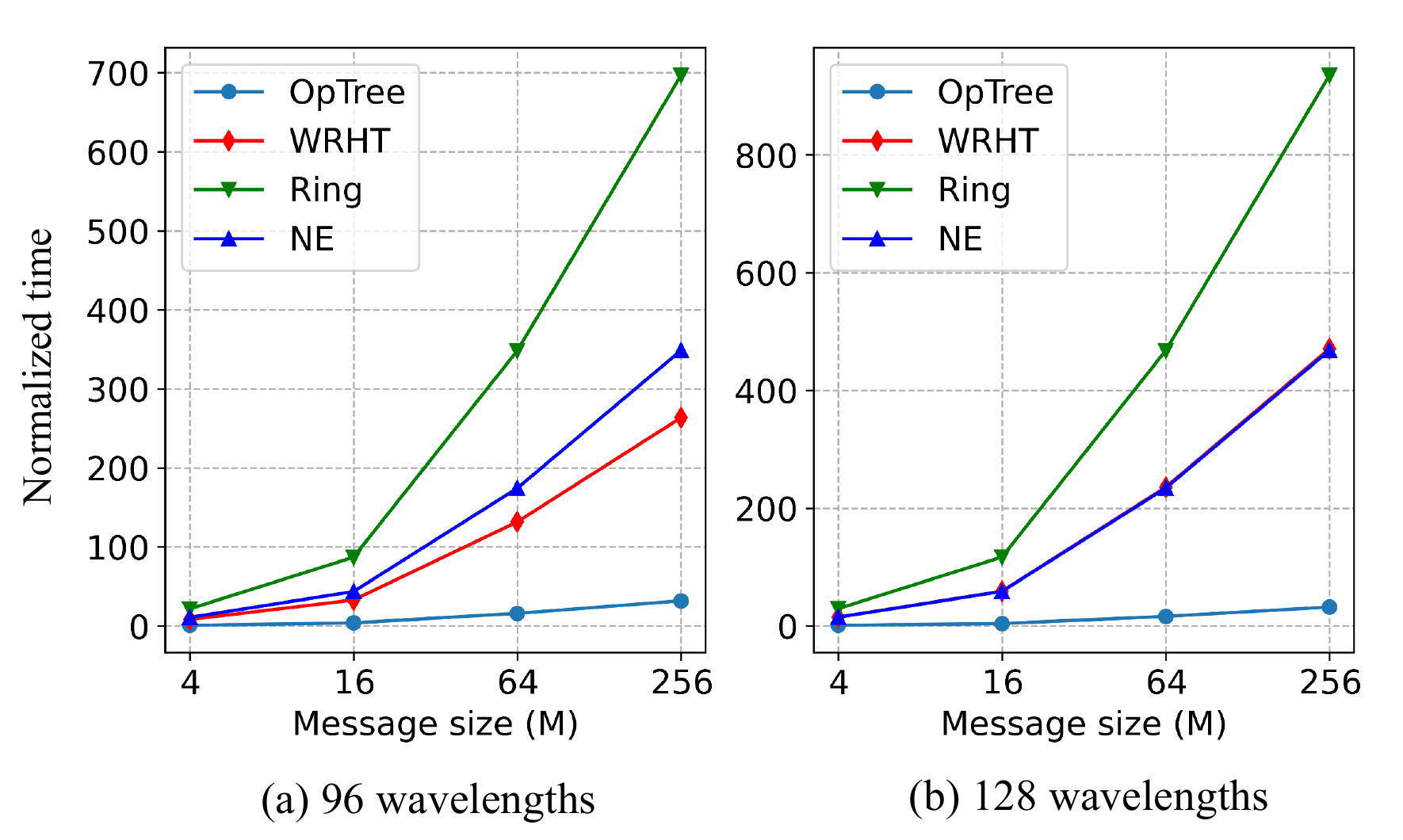}
		{
			\vspace{-7mm}
			\caption{Performance comparison of different All-gather algorithms in 1024-node optical interconnect system with different wavelengths}
			\label{fig:wavelengths}
		}
	\end{figure}

	In this set of simulations, we first compare the performance of OpTree with existing schemes including W\textsc{rht}~\cite{dai2022wrht}, Ring~\cite{chen2005performance}, and NE~\cite{chen2005performance} under 1024 and 2048 nodes in the optical interconnect system with message sizes ranging from 4M to 128M with 64 wavelengths. Then, we further compare these algorithms by increasing the available wavelengths to 96 and 128 under 1024 nodes. %We set the range of small messages from 32 to 1024K and the range of large messages from 4 to 4M. 
	All results in Fig.~\ref{fig:nodes} and Fig.~\ref{fig:wavelengths} are normalized by dividing the first result of OpTree.

	Fig.~\ref{fig:nodes} shows the performance comparison between OpTree, W\textsc{rht}, Ring, and NE algorithms for All-gather operation using different numbers of nodes. From Fig.~\ref{fig:nodes} (a) and (b), we can see that the time of OpTree is the lowest and the time of Ring is the highest among different message sizes. The time of these algorithms increases slowly with the increasing of message size. As the number of nodes increases from 1024 to 2048, the total time of W\textsc{rht} gradually reduces, getting close to OpTree. 
	That is because, with the increasing number of nodes, the number of communication steps of OpTree grows faster than that of W\textsc{rht}. Fig.~\ref{fig:wavelengths} shows the performance comparison between OpTree, W\textsc{rht}, Ring, and NE algorithms for All-gather operation using different wavelengths. From Fig.~\ref{fig:wavelengths} (a) and (b), we can see that OpTree has the least time cost and Ring has the
	highest time cost under both 96 and 128 wavelengths. When the wavelength increases from 96 to 128, the time for OpTree slightly decreases while W\textsc{rht} shows an increasing trend with the increase of messages size. Specifically, the time of WRHT is smaller than Ring but slightly larger than NE when the wavelength is set to 128.
	
	Compared to W\textsc{rht}, Ring, and NE, OpTree can reduce communication time by 56.36\%, 92.76\%, and 85.54\% on average in the optical interconnect system under different numbers of nodes, and can reduce communication time by 88.06\%, 95.84\%, and 91.69\% on average in the 1024-node optical interconnect system under different wavelengths, respectively. 
	
	%For large messages, the OpTree can reduce communication time by *\%, *\%, and *\% on average, respectively, compared with W\textsc{rht}, Ring, and NE. 
	%\begin{figure}[!ht]
	%	\vspace{-3mm}
	%	\centering
	%	\includegraphics[scale=0.55]{messages.pdf}
	%	{
	%		\vspace{-3mm}
	%		\caption{ Performance comparison of different All-gather algorithms in 1024-node optical interconnect system}
	%		\label{fig:data}
	%	}
	%\end{figure}
	
	\vspace{-1mm}
	\section{Conclusion} \label{sec:conclusion}
	%In this paper, we propose an efficient all-reduce scheme called W\textsc{rht} for distributed data-parallel DNN training in optical interconnect system to reduce the communication time. By reusing the wavelengths, W\textsc{rht} can obtain the minimum number of communication steps and optimal communication time for distributed DNN training. 
	In this paper, we propose an efficient scheme called OpTree for All-gather operation in optical interconnect systems. Based on OpTree, we derive the optimal $m$-ary tree corresponding to the optimal number of communication stages achieving the minimum communication time among all the OpTree options. 
	%by giving the number of nodes and wavelengths in optical interconnection systems. 
	We further analyze and compare the communication steps of OpTree with existing All-gather algorithms. Theoretical results demonstrate that OpTree requires much less number of communication steps than existing methods for All-gather operation on optical interconnect systems. %The constraint on the optical power of OpTree is also discussed.
	Simulations results show the effectiveness of OpTree, which can reduce communication time by 72.21\%, 94.30\%, and 88.58\% on average in the simulated optical interconnect system compared with three existing All-gather algorithms, respectively. %Simulation results show that OpTree can reduce the communication time by 72.97\%, 93.15\%, and 86.32\% on average under different message sizes compared with three existing All-gather algorithms simulated in the 1024-node optical interconnect system. Results also show that OpTree outperforms other All-gather algorithms under different nodes, particularly when the number of nodes is between 512 and 1024. Future work can be done by extending to other interconnect topologies and heterogeneous computing device scenarios.
	
	\bibliographystyle{IEEEtran}
	\bibliography{paper}
	
	%\appendix
	%\chapter{Proof}
	%this is the proof of theorem 1
	
\end{document}